\documentstyle[prl,multicol,aps,graphicx]{revtex}

\begin{document}

\title{Strong damping of phononic heat current by  magnetic
excitations in SrCu$_2$(BO$_3$)$_2$}

\author{M. Hofmann$^1$, T. Lorenz$^1$, G. S. Uhrig$^2$, H. Kierspel$^1$,
O. Zabara$^1$, A. Freimuth$^1$, H. Kageyama$^3$, and Y. Ueda$^3$}
\address{$^1$II. Physikalisches Institut, Universit\"{a}t zu K\"{o}ln, 50937 K\"{o}ln, Germany\\
         $^2$Institut f\"{u}r Theoretische Physik, Universit\"{a}t zu K\"{o}ln, 50937 K\"{o}ln, Germany\\
 $^3$Institute for Solid State Physics, University of Tokyo, Kashiwanoha 5-1-5, Kashiwa,
Chiba 277-8581, Japan}

\date{May 8, 2001}
\maketitle

\begin{abstract}
Measurements of the thermal conductivity as a function of temperature and
magnetic field in the 2D dimer spin system SrCu$_2$(BO$_3$)$_2$ are presented.
In zero magnetic field the thermal conductivity along and perpendicular to the
magnetic planes shows a pronounced double-peak structure as a function of
temperature. The low-temperature maximum is drastically suppressed with
increasing magnetic field. Our quantitative analysis reveals that the heat
current is due to phonons and that the double-peak structure arises from
pronounced resonant scattering of phonons by magnetic excitations.
\end{abstract}

\begin{multicols}{2}

During the last few years the thermal conductivity ($\kappa$) of low
dimensional spin systems has attracted considerable interest
\cite{nakam91,cohn95,baber98,ando98,vasil98,solog00}. One  reason is that in
these materials a large magnetic contribution $\kappa^{mag}$ to the heat
current may be present as observed e.g.\ for the spin ladder material
Sr$_{14-x}$Ca$_x$Cu$_{24}$O$_{41}$~\cite{solog00}. Another reason is that the
phononic heat current $\kappa^{ph}$ probes the spectrum of magnetic
excitations as well as the spin-phonon coupling~\cite{baber98,ando98,vasil98}.
The latter is very important in some of the low dimensional spin systems, e.g.\
in the spin-Peierls compound CuGeO$_3$~\cite{ando98,vasil98}. Both issues, the
magnetic contribution to the heat current as well as the interaction of the
phonons with magnetic excitations, are to a large extent unexplored and not
well understood.

A material of particular interest in this context is SrCu$_2$(BO$_3$)$_2$
(SCBO)~\cite{kagey99}. The Cu$^{2+}$ ions form a quasi-2D spin system which is
by virtue of the crystal geometry an experimental realization of the
Shastry-Sutherland model~\cite{shast81b}. In SCBO the intra-dimer and
inter-dimer couplings are of magnitude $J_1 \simeq 72$~K and $J_2 \simeq
43$~K, i.e. $J_2/J_1 \simeq 0.6$~\cite{knett00b}. As expected for this ratio
SCBO has a dimerized singlet ground-state separated from the excited triplet
states by a finite gap $\Delta \simeq 35$~K as seen in the magnetic
susceptibility~\cite{kagey99b} or in inelastic neutron
scattering~\cite{kagey00a}. From the latter it is also known that the triplet
excitations are almost dispersionless, i.e the group velocity is very small,
in agreement with theoretical calculations
\cite{miyah99,weiho99a,mulle00a,knett00e}. Thus, a sizeable $\kappa^{mag}$ is
{\em not} expected for SCBO making this material a natural
 candidate to study the influence of magnetic excitations on
$\kappa^{ph}$.

In this letter we present measurements of the thermal conductivity of SCBO
along ($\kappa_a$) and perpendicular ($\kappa_c$) to the 2D spin planes in a
large temperature (2.5-275~K) and magnetic field range (0-17~T). In zero
field, both $\kappa_a$ and $\kappa_c$ show pronounced double-peak structures
as a function of $T$. For both directions the low-$T$ maximum is drastically
suppressed by a magnetic field. We present a model based on a purely phononic
thermal conductivity and explain the double-peak structure and its field
dependence by strong damping of the phononic heat current due to resonant
scattering of phonons by magnetic excitations. With the same parameters our
model considerably improves the description of the magnetic field dependence
of the specific heat of SCBO~\cite{kagey00c}.

For our study two samples of rectangular-bar-shaped form of $\sim 0.6 \times
1.9 \times 3$~mm$^3$ with the long direction along the $a$ and $c$~axis,
respectively, were cut from larger single crystals  of SCBO grown by the
traveling solvent floating zone method~\cite{kagey00d}. The thermal
conductivity was measured with a conventional steady state method using
differential Chromel-Au +0.07\%~Fe thermocouples calibrated in magnetic
fields. Typical temperature gradients were of the order of 0.2~K. The absolute
accuracy of our measurements is of order $\pm 10 \%$ because of uncertainties
in the sample geometry whereas the relative accuracy is about one order of
magnitude better~\cite{note1}.

\begin{figure}
\begin{center}
    \includegraphics[width=7.8cm,clip]{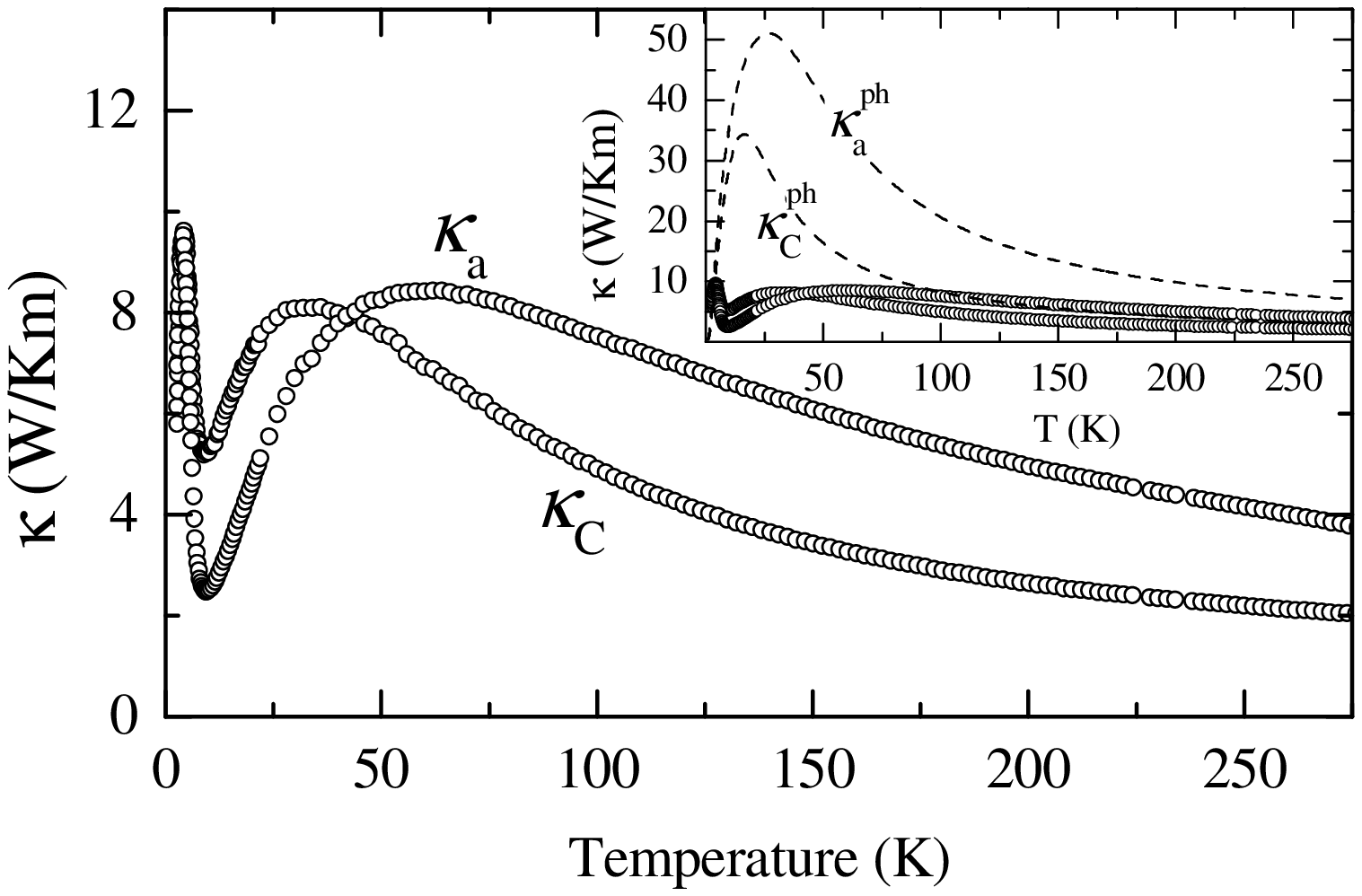}
\end{center}
\caption{Thermal conductivity of SrCu$_2$(BO$_3$)$_2$ along ($\kappa_a$) and
perpendicular ($\kappa_c$) to the magnetic planes. The inset shows the
experimental data ($\circ$) together with  ''conventional'' phononic
contributions $\kappa_a^{ph}$ and $\kappa_c^{ph}$ (-~-) obtained by switching
off the scattering of phonons by magnetic excitations (see text).} \label{fig1}
\end{figure}

\begin{figure}
\begin{center}
     \includegraphics[width=7.8cm,clip]{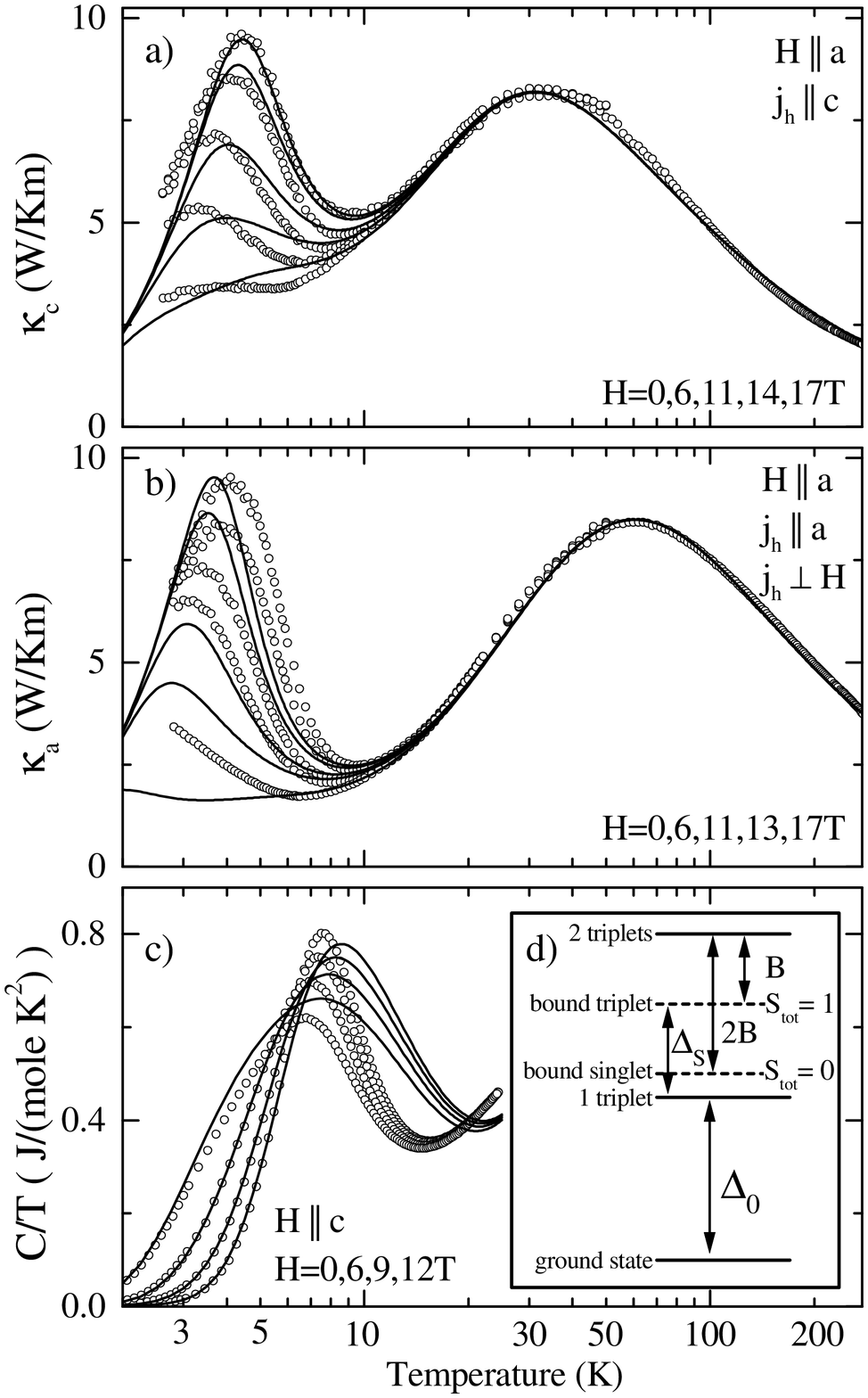}
\end{center}
\caption{Thermal conductivity $\kappa_c$ (a) and $\kappa_a$ (b) on a
logarithmic temperature scale for various magnetic fields (see figure). On
increasing field the low-$T$ maximum is suppressed. The lines are theoretical
curves calculated for the same magnetic fields via Eq.~(\ref{kphonon}). (c)
Specific heat at various fields  $H||c$ (the peak height decreases with
increasing field). Data points are from Ref.\ \protect\onlinecite{kagey00c};
the lines result from our theory. (d) Level scheme used in our model: The
solid lines denote the energies of the ground state, 1 triplet and 2
elementary triplets, respectively. The dashed lines indicate bound states with
$S_{tot}$ = 0 or 1, which are built from two elementary triplets. $B$ is the
binding energy and $\Delta_0$ is the elementary gap. $\Delta_s$ is the energy
difference relevant for the resonance scattering according to
Eq.~(\ref{rs-simple}).} \label{fig2}
\end{figure}

We show in Fig.~\ref{fig1} the thermal conductivity $\kappa_a(T)$ (= $\kappa_b$
in tetragonal SCBO) and $\kappa_c(T)$ in zero magnetic field.
For both directions pronounced double-peak structures are observed with the
low $T$ maxima occurring at $\approx 4.5$~K. For $\kappa_a$ the high $T$
maximum lies at $\approx 60$~K and for $\kappa_c$ at $\approx 30$K. Above
100~K we find $\kappa_c \propto T^{-1}$ as expected for phonon heat
transport~\cite{berma76} whereas $\kappa_a$ follows a $T^{-0.6}$ dependence.
The magnetic field dependence of the thermal conductivity is shown in
Fig.~\ref{fig2}. The low $T$ maximum is suppressed strongly by a magnetic
field for both $\kappa_a$ and $\kappa_c$. The magnetic field dependence at
higher temperatures is only weak ($\le 2 \% $) and close to the relative
measurement accuracy. The minimum of the double-peak structure and the low $T$
maximum systematically shift to lower temperatures with increasing magnetic
field.

The behavior of $\kappa $ observed here is reminiscent of that found in other
low dimensional spin systems~\cite{nakam91,ando98,solog00}. In particular, in
CuGeO$_3$ $\kappa$ also shows an almost field-independent maximum at $\sim
20$~K and a second one around $5$~K, which is strongly suppressed by a
magnetic field. This has been interpreted in terms of a magnetic

\noindent
and a phononic
contribution to the heat current with maxima at different
temperatures~\cite{ando98}.

For several reasons, a similar interpretation is not possible in
SCBO. Firstly, due to their flat dispersion the single triplet
excitations are not expected to contribute significantly to the heat current.
Secondly, $\kappa^{mag}$ in a 2D magnetic system should be strongly
anisotropic, as is observed e.g.\ in
Sr$_{14-x}$Ca$_x$Cu$_{24}$O$_{41}$~\cite{solog00}, but not in our
measurements. The lack of a strong anisotropy excludes also an explanation in
terms of multi-triplet excitations observed in SCBO above $\sim
5$~meV~\cite{kagey00a,kagey00c,nojir99,lemme00a}. These excitations move much
more easily than the single  triplet excitations\cite{knett00b}, but again
only within the magnetic planes. Finally, the relation $\kappa \propto c v
\ell $ ($c$ is the specific heat, $v$ the group velocity and $\ell $ the mean
free path) predicts similar $T$ dependencies of $\kappa^{mag}$ and the
magnetic specific heat $c^{mag}$. But measurements of $c$ in fields up to
12~T~\cite{kagey00c} show that $c^{mag}$ gives rise to a {\em maximum} of
$c/T$ at temperatures close to the {\em minima} of $\kappa_a$ and $\kappa_c$
for the respective magnetic fields (see Fig.~\ref{fig2}).

From these arguments we exclude a sizable magnetic contribution to the heat
current~\cite{note0}, so that $\kappa \approx \kappa^{ph}$. As the cause of
the double peak structure and its magnetic field dependence we propose
scattering of phonons by magnetic excitations. The absence of a strong
anisotropy of $\kappa_a$ and $\kappa_c$ is then easily understood, since both,
the anisotropy of the spin-phonon coupling and/or of the phonon system itself
are expected to be much weaker than that of the 2D magnetic system. In
addition, this scenario yields a straightforward explanation why the minima of
$\kappa_a$ and $\kappa_c$ occur at the maxima of $c/T$. The latter is the
temperature derivative of the magnetic entropy and thus directly related to
the number of magnetic excitations which serve as scatterers for the phonons.

For a quantitative description we fit our data by a Debye model for the
phononic thermal conductivity~\cite{neelm72}
\begin{equation}
\kappa^{ph} = \frac{k_{B}}{2 {\pi}^2
  v}\left(\frac{k_{B}}{\hbar}\right)^3 T^3
\int_{0}^{\Theta_{D}/T}
\frac{x^4 e^x\tau\left(\omega,T\right)}{\left( e^x - 1 \right)^2}dx \ ,
\label{kphonon}
\end{equation}
where $v$ is the sound velocity, $\Theta_{D}$ the Debye temperature, $\omega$
the phonon frequency, $x = \hbar\omega/k_{B}T$ and $\tau\left(\omega,T\right)$
the mean lifetime of a phonon. The total scattering rate $\tau^{-1}$ is given
by
\begin{equation}
{\tau}^{-1} = \tau_{bd}^{-1} +  \tau_{pt}^{-1} + \tau_{um}^{-1} + \tau_{res}^{-1} \;\; .
\label{rates}
\end{equation}
Here $\tau_{bd}^{-1}$, $\tau_{pt}^{-1}$, and $\tau_{um}^{-1}$ refer to the
standard relaxation processes for conventional phonon heat transport, i.e. to
boundary scattering, scattering by point defects, and umklapp scattering,
respectively~\cite{klemm58}. $\tau_{res}^{-1}$ describes the scattering of
phonons on magnetic excitations, which will be considered in the following.

For the double-peak structure of the phononic thermal conductivity we need a
scattering mechanism that is most active in a certain temperature range. We
therefore consider resonant scattering~\cite{neelm72,round70,toomb73,note2}.
The physical picture is the following: A phonon of energy $\omega$ is
absorbed, thereby exciting the magnetic system (under the constraint of spin
conservation). Then another phonon of the same energy is emitted and the
magnetic system is deexcited. If excitation (or deexcitation) starts from a
localized state like the dispersionless triplet excitations of SCBO there is
no directional correlation between the absorbed and the emitted phonon. The
net effect is therefore the same as that of point-like defects except for the
frequency dependence of the matrix element $M_{k',k}(\omega)$, and the
scattering rate $\tau_{res}^{-1}$ is of the form
\begin{equation}
\label{rs-simple}
\tau^{-1}_{res}(\omega) = C \omega^2 M_{k',k}^2(\omega ) F(T)
                        = C \frac{\omega^4 \Delta_s^2}{(\omega^2-\Delta_s^2)^2} F(T) \ .
\end{equation}
Here, $\Delta_s$ is the energy splitting of the two levels involved, $C$ is
the overall coupling strength between phonons and the two-level system and
$F(T)$ contains information on the state of the two-level
system~\cite{neelm72,round70,toomb73}. Notably, resonant scattering starting
from the magnetic ground state is not effective! The reason is that the ground
state is a coherent, translational invariant state with zero momentum so that
the emitted and absorbed phonons must have the same momentum and energy.

With this scenario in mind we consider resonant scattering within the magnetic
excitation spectrum of SCBO (Fig.~\ref{fig2}d). The lowest excited state is a
single triplet with excitation energy $\Delta_0$. It is thermally excited with
probability $c_1 = 1 - [1+ m_1 \exp(-\beta \Delta_0)]^{-1}$. Here $\beta =
(k_BT)^{-1}$ and $m_1 = 1 + 2 \cosh(\beta h)$ contains the magnetic field
dependence through $h= g\mu_BH$. In the resonance process a second triplet is
excited on one of the two adjacent dimers, and it combines with the first,
thermally excited triplet to a bound triplet state so that the total spin is
conserved. The energy of the bound triplet is given by $2 \Delta_0 - B$, where
$B$ is the binding energy~\cite{note4}. The resonance energy to be used in
Eq.~(\ref{rs-simple}) is therefore $\Delta_s = (2 \Delta_0 - B) - \Delta_0 =
\Delta_0 - B$. A complete description of the thermal factors of the resonance
process takes account of (i) the fact that a thermally excited triplet has to
be present and (ii) of all  resonance processes involving the 2 adjacent
dimers. This yields~\cite{note1}
\begin{equation}
\label{F1}
F(T)= 2 c_1\left(N_0+N_1 + (N_0-N_1)^2\right)
\end{equation}
with  the probabilities $N_0=1-c_1$ of having a dimer in its singlet state and
$N_1=c_1/3$ of having a dimer in the  triplet state with a particular $S^z$
component. Note that in the resonance process the total $S^z$ component is
conserved due to spin conservation. Therefore $\Delta_s$ does not depend on the
magnetic field. It is only $c_1$ that is field dependent.

In zero field Eqs.~(\ref{rs-simple},\ref{F1}) yield an excellent fit (not
shown) for $\Delta_0\approx 35$~K and $\Delta_s\approx 20$~K indicating strong
binding effects ($B\approx 15$~K). This is in good agreement with inelastic
neutron scattering and ESR experiments showing the lowest bound triplets at
$\approx 55$~K and an elementary gap of $\Delta_0 \approx 35$~K (see
Fig.~\ref{fig2}d)~\cite{kagey00a,nojir99}. However, the decrease of the thermal
conductivity in magnetic fields computed by Eqs.~(\ref{rs-simple},\ref{F1}) is
much stronger than that found in our experiments.

We can account for this discrepancy by considering a second, magnetic field
independent contribution to the resonance scattering. It arises from the
presence of bound {\em singlet} excitations with energy $2\Delta_0 -
2B$~\cite{knett00b,note4} (see Fig.~\ref{fig2}d). Since $2B \approx \Delta_0$
in SCBO, the energy of these bound singlets is close to that of single triplet
excitations yielding comparable thermal populations of these states. Therefore
bound singlets have to be treated as scattering centers for resonance
scattering too. In this case a dimer adjacent to a bound singlet is excited
and the corresponding 3 triplets combine to $S_{tot}=0$. Since $S_{tot}=0$,
this contribution to the resonance scattering does not depend on a magnetic
field and lowers the sensitivity of the total resonance scattering rate to a
magnetic field. The contribution of the bound singlets to the scattering rate
is treated in analogy to that of the single triplets. The probability that a
pair of adjacent dimers is in a state made of two triplets is given by $c_2 =
1-(1 + 2 m_1 \exp(-\beta\Delta_0))/Z$, where $Z$ is the complete partition sum
of the system of two triplets. In this way we achieve a theoretical
description comprising single triplets with density per dimer $n_1$ and
2-triplet states with density per dimer $n_2$. Exclusion effects - a dimer
cannot be in a single triplet state and involved in one of the bound states
simultaneously - are accounted for by $n_1 = c_1(1-4n_1-2n_2)$ and $n_2 =
2c_2(1-2n_1-3n_2)$.

A complication arises from a repulsive interaction $v$ between neighboring
triplets. The energy $\Delta_0$ of a single triplet is smaller than $J_1$ due
to virtual processes between neighboring dimers resulting from the dimer-dimer
coupling. If a triplet is excited on a neighboring dimer these processes are
(in part) blocked and $\Delta_0$ increases. Accounting for this on a
mean-field level yields $v = R(n_1+3n_2/2)$ with
$R=(J_1-\Delta_0)/2$~\cite{note1}. In the calculation of $\kappa $ one has to
replace $\Delta_0 \rightarrow \Delta_0 + p v$ in all equations (i.e. in $c_1$,
$c_2$ and the energies given in \cite{note4}), where $p=2$ refers to single
triplets and $p=3/2$ to bound states from two triplets.

To illustrate the validity of our model we compute the specific heat for
various magnetic fields. For the realistic parameters $\Delta_0=36$~K,
$R=18$~K (from $J_1=72$~K \cite{knett00b}) and $B=17$~K we obtain the results
depicted in Fig.~\ref{fig2}c. At low $T$ the agreement is very good;  position
and height of the peak are considerably improved relative to the isolated
dimer model \cite{kagey00c}. At higher $T$ we presume that the local model
developed above is too simplistic since the dispersion
\cite{knett00b,kagey00a} of the bound states is neglected.

The resonant scattering, decisive for the thermal conductivity, is calculated
in the {\em same} model with the {\em same} parameters. We obtain
$\tau^{-1}_{res} = C\omega^4(A_1+A_2)$ with
\begin{mathletters}
\begin{eqnarray}
\label{aeins}
A_1&=&
\frac{2n_1(N_0+N_1+(N_0-N_1)^2)(\Delta_0-B)^2}{((\Delta_0-B)^2-\omega^2)^2}\\
\label{azwei}
A_2&=&
\frac{3n_2(N_0+N_1+2(N_0-N_1)^2)\Delta^2_0}{(\Delta^2_0-\omega^2)^2}
\end{eqnarray}
\end{mathletters}
where $N_1=(n_1+3n_2/2)/3$ and $N_0=1-3N_1$. Using this $\tau^{-1}_{res}$
together with the usual scattering rates (Eq.~(\ref{rates})) we fit the
experimental $\kappa_a$ and $\kappa_c$ by Eq.~(\ref{kphonon}). The sound
velocity ($v=7600$~m/s) along the a-direction and $\Theta_{D} = 453$~K are
calculated from the measured elastic constant $c_{11}$~\cite{zherl00}. Due to
the lack of experimental data for $c_{33}$ we take the same velocity for the
fit of $\kappa_c$. The relaxation rate $\tau_{bd}^{-1}$ is then obtained by
$\tau_{bd}^{-1} = v/L$ with the characteristic sample lengths $L=0.75$~mm
($\kappa_a$) and 0.5~mm ($\kappa_c$) of the two samples. The point defect
scattering rate is given by $\tau_{pt}^{-1} = P{\omega}^4$ and the phonon
umklapp scattering rate is approximated by $\tau_{um}^{-1} =
UT{\omega}^{3}\exp(\Theta_{D}/uT)$~\cite{klemm58}. Since the  gap $\Delta_0$,
the binding energy $B$, and the interaction $R$ are determined from  the
specific heat there remain only four adjustable parameters ($P$, $U$, $u$, and
$C$).

As shown in Fig.~\ref{fig2} our model yields an almost perfect fit of the
measured $\kappa_c$ (for $H=0$) over the entire $T$ range. Above 10~K the fit
of $\kappa_a$ is also almost perfect, whereas below 10~K a slight $T$ shift
($\le 1$~K) between fit and experimental data occurs. We use $P=7.7\cdot
10^{-43}$~s$^3$ ($7.9\cdot 10^{-43}$~s$^3$), $U=2.6\cdot 10^{-31}$~s$^2$/K
($5.7\cdot 10^{-31}$~s$^2$/K), $u=5$ (10), and $C=3.35\cdot 10^{6}$~s$^3$
$(1.15\cdot 10^{6}$~s$^3$) for the calculation of $\kappa_a$ ($\kappa_c$).
These values are reasonable for an insulator and are comparable in magnitude
to those found e.g.\ in the spin ladders~\cite{solog00}. Different values for
$\kappa_a$ and $\kappa_c$ are expected in an anisotropic crystal.

Accounting for the Zeeman splitting in $c_1$ and $c_2$, we calculate
$\kappa_a$ and $\kappa_c$ in magnetic fields without further parameter
adjustment. Using $g\approx 2.07$ \cite{kagey99b}, our model reproduces the
overall influence of the magnetic field very well. The high $T$ maxima remain
unchanged whereas the low $T$ maxima are continuously suppressed. Again the
agreement between calculated and experimental values is better for  $\kappa_c$
than for $\kappa_a$. Whereas the field influence on $\kappa_a$ is slightly
overestimated by our model the experimental values of $\kappa_c$ are even
quantitatively reproduced up to the highest field.

In the calculations, one can switch off the resonant scattering on magnetic
excitations by setting $C=0$. $\kappa^{ph}(C=0)$ obtained in this way is much
larger than the measured thermal conductivity, as shown in the inset of
Fig.~\ref{fig1}. As expected for resonant scattering the damping of the phonon
heat transport is most pronounced for $T \approx \Delta_s $. Note, however,
that even at room temperature the suppression is sizeable. This strong damping
gives further evidence for a large spin phonon coupling in SCBO as has  been
inferred before from sound wave anomalies~\cite{zherl00}.

In summary, the thermal conductivity of the low dimensional quantum spin
system SrCu$_2$(BO$_3$)$_2$ has a characteristic double-peak structure with
two maxima at low temperatures and a pronounced magnetic field dependence. A
significant magnetic contribution to the heat current can be excluded, since
the double peak structure is not anisotropic and because the magnetic
excitations are (almost) localized. Our quantitative analysis in terms of
resonant scattering of phonons on magnetic excitations explains the double
peak structures and their magnetic field dependence very well and gives
evidence for strong spin-phonon coupling. Therefore, we call for further
investigations whether the same mechanism is at work in other low-dimensional
spin systems.

We acknowledge useful discussions with E.~M\"{u}ller-Hartmann, M.~Gr\"{u}ninger and
A.~P.~Kampf. This work was supported by the Deutsche Forschungsgemeinschaft in
SFB 341 and  SP 1073.

\end{multicols}

\begin{thebibliography}{10}
\bibitem{note0}
A small magnetic contribution to $\kappa_a$ can of course not be excluded.
Such a contribution might explain in part the different temperature
dependences of $\kappa_a$ and $\kappa_c$, but it is unimportant in the context
of the double peak structure.
\bibitem{note1}
Details will be presented in a forthcoming publication.
\bibitem{note2}
The more conventional scattering mechanisms where e.g. a phonon decays into
two magnetic excitations are far less effective, basically because of
the discreteness of the magnetic excitation spectrum.
\end{thebibliography}
\end{document}